# Cancer Cell Motility: Optimizing Spatial Search Strategies

L. Leon Chen, Le Zhang, Jeongah Yoon, and Thomas S. Deisboeck

Complex Biosystems Modeling Laboratory, Harvard-MIT (HST) Athinoula A. Martinos Center for Biomedical Imaging, Massachusetts General Hospital, Charlestown, MA 02129, USA.

**Running Title:**     Optimizing Spatial Search Strategies in Cancer

# Corresponding Author:

Thomas S. Deisboeck, M.D.
Complex Biosystems Modeling Laboratory
Harvard-MIT (HST) Athinoula A. Martinos Center for Biomedical Imaging
Massachusetts General Hospital-East, 2301
Bldg. 149, 13th Street
Charlestown, MA 02129
Tel:    617-724-1845
Fax:    617-726-7422
Email: deisboec@helix.mgh.harvard.edu

# ABSTRACT

Aberrantly regulated cell motility is a hallmark of cancer cells. A hybrid agent-based model has been developed to investigate the synergistic and antagonistic cell motility-impacting effects of three microenvironment variables simultaneously: chemoattraction, haptotactic permission, and biomechanical constraint or resistance. Reflecting distinct cell-specific intracellular machinery, the cancer cells are modelled as processing a variety of spatial search strategies that respond to these three influencing factors with differential weights attached to each. While responding exclusively to chemoattraction optimizes cell displacement effectiveness, incorporating permission and resistance components becomes increasingly important with greater distance to the chemoattractant source and/or after reducing the ligand's effective diffusion coefficient. Extending this to a heterogeneous population of cells shows that displacement effectiveness increases with clonal diversity as characterized by the Shannon index. However, the resulting data can be fit best to an exponential function, suggesting that there is a level of population heterogeneity beyond which its added value to the cancer system becomes minimal as directionality ceases to increase. Possible experimental extensions and potential clinical implications are discussed.

# 1. INTRODUCTION

Rapid and extensive cell migration is a central component of what drives local cancer invasion and distant metastasis (Friedl and Wolf, 2003; Yamaguchi et al., 2005). This ability to relocate is conferred by the gain of motility capabilities and the loss of various regulatory mechanisms (Condeelis et al., 2005). While this originates at the genetic level, interactions with the microenvironment are crucial in the expression of this phenotype, a phenotype that includes a spectrum of behaviors and responses (Friedl and Wolf, 2003).

While there may be a variety of factors that influence cancer cell migration, three distinct mechanisms are considered: chemotaxis ($A$), permission ($P$), and resistance ($R$). Chemotactic *attraction* is the directed motion along a concentration gradient of diffusive ligands. These gradients induce bias in what would otherwise be a random walk, the strength of which affects the overall persistence of the chemotaxing cells (Arrieumerlou and Meyer, 2005). While chemotaxis is a large component of directional migration, it is known that substrate-bound ligands also have significant influence on migration by mediating the detachment force of adhesive receptors such as integrins (Hood and Cheresh, 2002). The speed at which cells migrate exhibits a biphasic behavior, with maxima occurring at different concentrations of extra-cellular matrix (Palecek et al., 1997). This is a function of the level of integrin expression, while the maximum speed attainable, however, remains constant (Galbraith and Sheetz, 1998). This effect will herein be referred to as haptotactic *permission*. A third microenvironmental factor is mechanical *resistance*. From a physical perspective, cancer cells will tend to be nudged or forced to regions of lower mechanical resistance, based on physical principles (Deisboeck et al., 2001; Wurzel et al., 2005; Zaman et al., 2005). As cells secrete enzymes (i.e., proteases) that degrade the surrounding extracellular matrix, pockets of lower resistance result. Here, the resistance layer of the microenvironment models this effect, which induces cells to follow one another through these trails of lower resistance. Indeed, this is not the only mechanism by which cancer cells may "cooperate" during invasion, but it captures an important component. Other mechanisms, for example, include clustered migration, paracrine loops of released signaling factors, and induction of adhesion molecules on stromal cells (Deisboeck and Couzin, 2008). It has been suggested that successive cancer cell migration conditions the microenvironment so that subsequent cells have an increasingly easier time invading the tissue (Bidard et al., 2008; Deisboeck et al., 2001). This is exactly what is captured by $R$, resistance.

While there have been various quantitative experimental studies performed on cancer cell migration, for example on chemotaxis or haptotaxis, there has not been any systematic study separating individual effects. Here, *in silico* modeling is poised to generate testable hypothesis in the face of a dearth of quantitative data. We describe in this paper a simple computation model of cancer cell migration within a microenvironment which incorporates these several key influencing factors.

# 2. METHODS

## 2.1 Simulation strategy

To simulate the cancer cell search decision process, a discrete-continuum *hybrid, agent-based model* was developed. This model attempts to investigate the contributions of environmental conditions that trigger chemotaxis, affect permission, and represent resistance. Here, we adopt the assumptions that these three are the only effects governing cell migration, and that all cells are strictly migratory (as opposed to also engaging in proliferation; we plan on relaxing this assumption in future works). Migratory behavior is modeled as a search process, where a given cell attempts to reach its goal within the constraints of its search decision paradigm. That goal is defined as positive movement towards the chemoattractant source while simultaneously minimizing energy expenditure.

The foundation of the discrete part of the model is a 100x100 square lattice representing the 2D microenvironment. The cells are placed upon this lattice grid and allowed to migrate on it as well as remodel it through various feedback mechanisms. Interactions occur with respect to the three tissue qualities.

*2.2 Search paradigm*

Each cell follows a search paradigm that governs how it arrives at its movement decisions. Any such search paradigm consists of three coefficients $[C_A, C_P, C_R]$ which serve as weights to each of the three aforementioned processes, describing the extent to which each contributes to the overall search decision. For example, a cell with a search paradigm of [1, 0, 0] would only allow chemotaxis to direct its migration, whereas a cell with search paradigm [0, 1, 0] would only allow permission to enter its decision process, while a cell with search paradigm [0, 0, 1] would only follow movement decisions made by taking biomechanical resistance information into account. Anything in between would allow proportional contributions from all three, but because these are relative weights, the following relation must hold:

$$C_A + C_P + C_R = 1 \tag{1}$$

*2.3 Microenvironmental setup*

The first tissue property is the chemoattractant field ($A$), which is currently comprised of glucose only. The chemotactic machinery within the cells responds to glucose gradients, creating bias in a probabilistic manner to what would otherwise be a random walk (Arrieumerlou and Meyer, 2005; Jabbarzadeh and Abrams, 2005). A point source of glucose is located at (75,75) and is constantly replenished, representing a blood vessel. Glucose from the source diffuses over the microenvironmental lattice on a timescale much smaller than the timescale of cell migration. A Crank-Nicolson algorithm (Smith, 1985) was implemented to solve, using Dirichlet boundary conditions, the following partial differential equation describing the diffusion mechanism:

$$\frac{\partial A}{\partial t} = D_{glu\cos e} \nabla^2 A, \tag{2}$$

where $A$ represents the concentration of the diffusible chemoattractant (i.e., glucose, in the current iteration). The glucose field is updated through the following discrete equations:

$$A_t(75,75) = S_{glu\cos e} \tag{3}$$

$$A_{t+1}(\mathbf{x},\mathbf{y}) = A_t(\mathbf{x},\mathbf{y})\Delta t_{cell}U_{glu\cos e} \tag{4}$$

where $\mathbf{x}$ and $\mathbf{y}$ are the components of the cell position vector, $S$ is the chemoattractant source, $U$ is the chemoattractant uptake rate per cell, and $\Delta t_{cell}$ is the time step for cell migration.

The second tissue quality, permission ($P$), remains a constant value throughout a simulation run. For all the simulation protocols except the series investigating the influence of a permission gradient at various locations on the lattice, it is kept at a uniform level. As we will see, this effectively serves as a source of randomness.

Finally, the third tissue quality of resistance ($R$) is updated according to the following equation:

$$R_{t+1}(\mathbf{x},\mathbf{y}) = R_t(\mathbf{x},\mathbf{y}) - \delta_{resis\tan ce} \tag{5}$$

where $\mathbf{x}$ and $\mathbf{y}$ are the components of the cell position vector, and $\delta_{resistance}$ is the relative reduction in confinement by (cell secreted) proteases.

*2.4 Cell decision process*

At each time step, cells process migration decisions based upon values of the three tissue qualities in their von Neumann neighborhoods. Values are weighted and converted to probabilities according to the following rules:

$$\Pr_A(x_i^{t+1},y_i^{t+1}) = \frac{\dfrac{A(x_i^t,y_i^t)-A(x_0,y_0)}{A(x_0,y_0)}}{\displaystyle\sum_{N^v_{(x_0,y_0)}}\dfrac{A(x_i^t,y_i^t)-A(x_0,y_0)}{A(x_0,y_0)}}, \quad i=1,2,3,4 \tag{6}$$

$$\Pr_P(x_i^{t+1},y_i^{t+1}) = \frac{P(x_i^t,y_i^t)\quad P(x_0,y_0)+\left|\min P(x_i^t,y_i^t)\right|}{\displaystyle\sum_{N^v_{(x_0,y_0)}}\left(P(x_i^t,y_i^t)\quad P(x_0,y_0)+\left|\min P(x_i^t,y_i^t)\right|\right)}, \quad i=1,2,3,4 \tag{7}$$

$$\Pr_R(x_i^{t+1},y_i^{t+1}) = \frac{R(x_i^t,y_i^t)}{\displaystyle\sum_{N^v_{(x_0,y_0)}}R(x_i^t,y_i^t)}, \quad i=1,2,3,4 \tag{8}$$

where $N^v_{(x_0,y_0)}$ are the locations in the von Neumann neighborhood about the current position ($x_0$, $y_0$). A higher probability value for a certain location means it offers more favorable conditions as evaluated by the subscript mechanism and thus, the cell is more likely to move to that location based on that mechanism alone.

We note that $A$, $P$, and $R$ all influence the cell simultaneously, however. The ultimate decision for migration is therefore determined by multiplying the corresponding weight to each of the three probabilities and summing:

$$\Pr(x_i^{t+1}, y_i^{t+1}) = C_A \Pr_A(x_i^{t+1}, y_i^{t+1}) + C_P \Pr_P(x_i^{t+1}, y_i^{t+1}) + C_R \Pr_R(x_i^{t+1}, y_i^{t+1}), \quad i = 1,2,3,4. \quad \textbf{(9)}$$

The above equation gives the probabilities for moving to each of the four lattice sites in the von Neumann neighborhood.

Both *in vitro* and *in vivo* observations have revealed diverse patterns of invasion into the local surrounding tissue. For instance, the tumor can shed individual cells, or disseminate via larger contractile units made of many cells (Friedl and Wolf, 2003). Studies have shown that individual invading cancer cells are the main cause of systemic dissemination and metastasis. While many tumors exhibit both forms of invasion, it is thought that individual cell migration is correlated with lower differentiation (Thiery and Morgan, 2004). Some highly differentiated tumors may in fact only invade through collective means (Friedl and Wolf, 2003). While in theory the mean displacement effectiveness of the population of cells may be also measured, here normalized displacement effectiveness is evaluated for the single most successful cell, and is defined as follows:

$$DE = \max \frac{\sum_{t=0}^{T} \left( \left\| x_{i,t}, y_{i,t} \right\| \quad \left\| x_{i,0}, y_{i,0} \right\| \right)}{\left\| x_{i,t}, y_{i,t} \right\| \quad \left\| x_{i,0}, y_{i,0} \right\|}, \quad \textbf{(10)}$$

where $x$ and $y$ are the positions of cell $i$ at time $t$. The displacement of the cell is normalized against the initial distance to the nutrient source, located at (75,75).

*2.5 Simulation protocols*

A series of *in silico* experimental protocols are performed to investigate the impact of various parameters and factors on the cell search strategy, both overall as well as optimally. As a baseline case, simulations with the following setup were performed (see **Figure 1**): $\lambda$=707.1 μm and $\alpha$=1, with a uniform permission distribution. $\lambda$=707.1 μm is achieved by fixing the nutrient source at (75,75) and initializing the cells in a 10x10 lattice about the point (50,50). 100 cancer cells are implemented and due to the fact that only one cell is allowed to occupy any given lattice site at a time, the initialization lattice must be 10x10 in dimension. $\alpha$=1 simply indicates that the baseline glucose diffusion coefficient is used. In addition, a uniform permission distribution simply adds a random walk component to cell motility, with the degree to which it is a factor being dependent on $C_P$. All model parameters used are listed in **Table 1**.

Changing the different parameters of the system will likely alter displacement effectiveness across the search paradigm space, in such a way that will lead to insights about cell motility that is not necessarily predetermined. Here, we investigate the effect of three factors: the initial location of the cells, the chemoattractant diffusion coefficient, and the introduction of a non-uniform permission gradient at various locations, thus reflecting a more realistic microenvironmental representation.

The displacement effectiveness at 24 hours of simulation time over the entire search paradigm space was determined for five initial locations: (50,50), (40,40), (30,30), (20,20), and (10,10). These coordinates are the center of the 10x10 initial lattice and correspond to distances to the chemoattractant source of 707.1, 989.9, 1273, 1556, and 1839 μm, respectively. (50,50) is the baseline case, and serves as the initialization location of all subsequent simulations.

Diffusion of ligands in extracellular spaces can be modeled by the modification of the standard diffusion equation with measures of tortuosity and the volume fraction (Nicholson, 2001). While this inherently takes into account the spatial heterogeneity of tissue, one could expect these to be dynamically changing over time as well. The result will be such that the effective diffusion coefficient of a certain diffusible ligand will be constantly fluctuating. To investigate the influence of the diffusion coefficient over a range, simulations are run with $D* = \alpha D$, where $\alpha$ is varied from 0.1 to 1.

As a non-diffusible tissue quality, permission has the capability of inducing a random walk component to cell motility when its distribution is uniform, or can function as another source for directional migration when there is a concentration gradient (Gunawan et al., 2006; Rhoads and Guan, 2007; Smith et al., 2006). In this iteration of our model, its distribution remains fixed throughout a simulation run, without a production or decay term. The influence of a permission gradient is investigated by distributing it non-uniformly:

$$P(x,y) = \exp\left(\frac{-((x - x_0)^2 + (y - y_0)^2)}{\sigma^2}\right), \tag{11}$$

where $(x_0, y_0)$ represents the location of the peak permission, and $\sigma$ controls the width. In a series of multiple simulations, $\sigma$ is arbitrarily set to 10 units, and $(x_0, y_0)$ is set to (0,0), (25,25), (50,50), or (75,75). This corresponds to a distance, from the initial position of the cells and in its relation to the direction to the source, of -1414, -707.1, 0, or 707.1 μm, respectively.

The independent effects of $A$, $P$, and $R$ is illustrated in **Figure 2**. $A$ confers directionality, so that the most successive and least successive cells operating by this information alone move nonetheless in the same direction towards the source. Cells operating using $P$ by itself exhibit random walks: the most successive cell and least successive cell have opposite directions. $R$ is a quality that only affects two or more interacting cells, and so cells that utilize only $R$ are stationary. In other words, only when $A$, $P$, or both are taken into consideration alongside $R$ does $R$'s effect emerge.

Finally, the genetic and/or epigenetic diversity of the cell population is varied by initializing subpopulations of cells to different search paradigms. The Shannon diversity index is used to quantify the degree of heterogeneity (Maley et al., 2006). In all previous simulations, the cells represented a homogeneous population with respect to their search paradigms, because for each run they were all initialized to a specific single search paradigm.

## 3. RESULTS

*3.1 Baseline case*

For each experimental setup, multiple simulations were performed to find the normalized displacement effectiveness (DE) over the entire search paradigm space. The result from the baseline setup is shown in **Figure 3**. For $C_P$=0, a sharp decrease in DE is observed with increasing $C_R$, whereas for $C_R$=0, the drop-off in DE is similar but slightly less dramatic. The maximum DE overall occurs for the coefficient weight combination of $C_A$=1, $C_P$=0, $C_R$=0. This reveals that for the particular microenvironment and initial conditions simulated, $P$ and $R$ do not aid in maximizing DE. Because there is a certain amount of stochasticity involved, this does not imply that a cell with such a search paradigm will always yield the maximal DE, but will rather do so on the average.

### 3.2 Varying the distance to the source

Displayed in **Figure 4** is the DE over the search paradigm space for four initial cell locations. When the cells are initialized close to the source, $C_A$ dominates. As the initial position of the cells is moved slightly away, $C_R$ begins making a contribution. This suggests that at medium distances where the chemotactic strength is moderate, cells following one another's preformed paths possibly become an important mechanism. At large distances, however, where the chemotactic strength is weak, $C_P$ begins to become significant as cells must undergo migration more akin to random walks to maximize DE.

Note, however, that the overall magnitude of the DE falls for increasing distances. Because DE is defined with respect to the source location over the same length of time for all cases, it is implicit that at greater distances, DE should be lower. This emphasizes the important notion that at greater distances, $C_P$ and $C_R$ do not necessarily absolutely increase DE, but rather become *relatively* important by virtue of the fact that $C_A$ becomes weaker. Intuitively, this is reasonable, as $C_A$ provides the only directional information, which decreases in strength as a cell moves farther away from the source.

### 3.3 Varying the diffusion coefficient

From the baseline case, we then varied the diffusion coefficient of the chemoattractant. **Figure 5** displays the DE over the search paradigm space for four diffusion coefficients. At lower diffusion coefficients, which can arise for example from higher tortuosities (Nicholson, 2001) or larger molecular sizes, overall DE decreases and the contribution of permission and resistance becomes increasingly more important.

The overall trend is quite similar to that of initial cell distance. For example, at intermediate diffusion coefficients, $C_R$ helps boost DE, perhaps more so than $C_A$ by itself. At low diffusion coefficients, $C_P$ becomes very important. There is, however, a notable difference in the level of DE at which the contributions from $C_P$ and $C_R$ begin to be more or less equal in importance. With increasing distance, that level of DE was around 0.1; with a decreasing diffusion coefficient, that level of DE was around 0.2. This implies that the distance to the chemoattractant source has a more pronounced effect on the magnitude of DE when contribution from $C_A$ is weak.

In **Figure 6**, DE is plotted against α over many more values of α for the search paradigm [$C_A$=1, $C_P$=0, $C_R$=0]. Note the rapid phase transition from low DE to high DE. In the region of high DE, chemotactic attraction predominantly maximizes DE while in the region of low DE, permission and resistance also play a large role in maximizing DE.

*3.4 Addition of permission gradient*

The results presented thus far have been with a uniform permission distribution. In other words, permission has been acting as a source of imprecision or stochasticity. Interestingly, but perhaps not altogether surprisingly, this randomness becomes an important contributor to maximizing DE in a variety of microenvironments. Specifically, we have demonstrated that it becomes proportionately more important as a cell moves further away from the chemoattractant source, and as the diffusion coefficient of the chemoattractant is reduced. At the other extreme, it is also possible for a non-diffusible ligand substrate to act as a source of constant directionality. This was modeled as a radial distribution with a peak at a certain location on the lattice. Next, we introduced this characteristic peak, and varied the location at which the peak was placed. Shown in **Figure 7** are plots of DE for different permission peak locations, where permission is distributed in such a way to create a gradient towards the peak. When the permission peak is located in the same direction as the chemoattractant source, a *synergistic* effect is seen, where search paradigms utilizing permission as well as chemotaxis information to direct cell migration maximizes DE. When the peak is located in the opposite direction as the source, an *antagonistic* effect is seen, with the strength of the antagonism having a slight dependency on distance for low $C_P$. This can be contrasted to the situation when uniform permission is present, where permission becomes less important in the search process with closer distances to the source or higher diffusion coefficients, and more important when farther away from the source or in the presence of lower diffusion coefficients.

*3.5 Impact of heterogeneity*

Finally, in **Figure 8**, DE is shown as a function of the genetic and/or epigenetic diversity of the initial cell population. The Shannon index is used as the measure of heterogeneity, and as one can conclude from the plot, cell heterogeneity (and thus search paradigm diversity) increases DE. Here, because the microevolutionary effects of increased heterogeneity (e.g., metapopulation competition) are not taken into consideration, the function is monotonic. The data is fitted to two functions. The linear fit to the equation $y = a*x + b$ yields a $R^2$ of 0.9288, while the exponential fit to the equation $y = a - b*\exp(-c*x)$ yields a $R^2$ of 0.9927, where $x$ is the Shannon index and $y$ is the displacement effectiveness. The exponential fit is more convincing since the first and last points fall on the best-fit line, and, for the reasons stated in the discussion section, it also appears more plausible from a biological perspective.

## 4. DISCUSSION

Mathematical and computational modeling of cell biology is increasingly being recognized as a valuable tool in elucidating the relative importance of underlying mechanisms. However, no computational model can capture all the intricacies of a behavior as complex as cell migration. This is perhaps even more true of cancer cell motility, as such behaviors result from a deregulated process, and can therefore be manifested in many more ways than one. Computational modeling then serves as a platform upon which further experimental work can be built. Such an iterative process is a central theme discussed in *systems biology* research (Fisher and Henzinger, 2007). Currently, most experiments focus on one feature by, in essence, deliberately overstating it, such as inducing chemotaxis through setting up strong diffusive gradients of a particular ligand, without being able to reproducibly and quantitatively alter the others. However, a better understanding of cell motility, and acquiring the ability to modify it, requires the knowledge of how these various factors interact with one another.

Here, we created as simple a model as was appropriate in attempts to gain new insight into relative contributions of three underlying forces of cell migration. Chemoattraction served as a source of directionality, and permission and resistance characteristics interacted in complex ways that generated interesting displacement effectiveness *landscapes*. These landscapes varied across different microenvironmental conditions, but trends were observed. Larger distances and lower diffusivities suppress the magnitude of displacement effectiveness achieved by chemoattraction alone, which effectively increases the relative role of permission and resistance. In addition, a permission gradient can contribute either synergistically or antagonistically to attraction, depending on the direction of the gradient. Under certain conditions, for example in **Figure 4c**, incorporation of a uniform permission distribution into a cell's search decision process (non-zero $C_p$) may actually optimize displacement effectiveness. A uniform permission distribution essentially acts as a source of randomness, and this result is supported by our previous work which has found that some level of randomness in search precision can in fact improve spatio-temporal expansion upon an entirely deterministic approach (Mansury and Deisboeck, 2003).

Interestingly, as diffusivity was varied, a rapid phase transition was observed from a low displacement effectiveness regime to a high displacement effectiveness one. It is not entirely clear why the transition is abrupt rather than gradual, as one would expect, and may simply be because the signal is not able to reach the cell within the experimental timeframe. This is a surprising result given that there are no explicit diffusivity thresholds within the model setup, and it would thus be interesting to see if such a phase transition actually existed in an experimental setting. One could potentially investigate this by using different chemoattractants with varying intrinsic diffusion coefficients, with the modulation of the uptake of these chemoattractants through genetic engineering of appropriate cell surface receptor systems (for the glucose transport receptor (GLUT), used in here as an example, this has in fact been attempted already (Kim et al., 2006)).

An important synthesis of all this information is the recognition that these displacement effectiveness landscapes are not of search paradigms per se, but of cells, each of which may adopt any possible search paradigm. In other words, these landscapes create different possibilities within a *population* of cells, which, together, can work in the favor of such heterogeneous populations. Clonal diversity allows higher probabilities of cells adopting search paradigms that yield higher displacement effectiveness. In terms of evolutionary selection, it can be expected that selection for fitter clones should occur faster than in less diverse populations. Numerous studies support this conclusion (Kuukasjarvi, 1997; Maley et al., 2006; Prest et al., 1999; Taniguchi et al., 1994). However, under certain microenvironmental conditions, when the displacement effectiveness landscape is flat, a less genetically and/or epigenetically diverse population of cells may not be any better or worse than a more heterogeneous one. Indeed, an exponential fit in **Figure 8** is observed because there may be diminishing returns associated with increasing heterogeneity (Solé and Deisboeck, 2004).

One of the ultimate goals of better understanding cancer cell motility and migration is to develop therapeutics that target, reduce, or ideally, halt this behavior in patients. Optimizing therapeutic efficacy requires an understanding of how the modification of an underlying process fits into the larger scheme of interacting processes, such as those involving $A$, $P$, and $R$. As an example, anti-angiogenesis therapies indirectly impact $A$ by affecting the chemoattractant source. While they may not be directly inhibiting the chemoattractant supply, but rather normalizing the vasculature that delivers it (Jain, 2008), they nonetheless should have the biggest impact on $C_A$-heavy cells close to a source. This suggests that anti-angiogenesis therapies may indirectly have an effect on invasion on certain subpopulations of cancer cells, and not just on tumor (re)growth. For similar reasons, drugs that disrupt downstream cell-signaling pathways of receptors processing chemotactic signals would be more efficacious closer to the source, meaning intravascular delivery may be actually the best method of delivery for these drugs.

Because the cellular response is different depending simultaneously on the search paradigm (intrinsic) and microenvironmental characteristics (extrinsic), based on our findings, one may want to therapeutically target the microenvironment as well. And indeed, this has been suggested in various other contexts (Albini and Sporn, 2007; Liotta and Kohn, 2001), but the case is made here in terms of microenvironmental characteristics such as permission and resistance. Those tissue qualities can be modified by inhibiting proteolytic remodeling of the extracellular matrix through tumor-secreted matrix metalloproteinases (Nakano et al., 1995; Overall and Kleifeld, 2006). Inhibitors of these have been in various clinical trials for years, but none have ever passed due to the preponderance or severity of side effects (Fingleton, 2007). Insights gained from this and similar studies can determine in what situations decreasing the effect of resistance may be more beneficial and which regions better be spared. We hypothesize that anti-proteolytic treatments may offer more benefit in areas distant to nutrient sources, where $P$ and $R$ become more important with respect to the cancer cell search decision process, thus making a case against vascular delivery of such agents. On the other hand, there are certain situations that may modify permission in an undesirable manner. A common side effect of radiation therapy is fibrosis, where the tissue becomes stiffer and thus possibly more permissive for migration (Stone et al., 2003). In this situation, cells that are $C_P$-heavy would be able to more effectively incorporate these permission gradients into its spatial search strategy.

While its potential clinical implications are intriguing, one has to remain cautious in extrapolating our findings just yet; our model is currently relatively simple, and thus it is unlikely that it has captured all aspects of motility. For example, one such possibly important factor that was not taken into consideration here was cell density. Cell density may modulate the effects of resistance (Couzin et al., 2005; Deisboeck and Couzin, 2008; Szabo et al., 2006). Cell proliferation could in theory be an important aspect of this, thus we plan to incorporate cell proliferation in a future iteration. The inclusion of cell proliferation and perhaps cell death would then also capture to a certain degree adaptation and the evolvability through the search paradigm space. All of this would be part of an effort to try and incorporate patient-specific microenvironmental lattices into *in silico* tumor modeling and simulation (Huang et al., 2008), which is motivated in part by growing advances in imaging (Diehn et al., 2008; Weissleder and Pittet, 2008).

In summary, despite such shortcomings, this model already offers some interesting new insights into a cancer cell's spatial search processes. In advancing anti-cancer treatment strategies, the biological characterization of different components of the microenvironment and how they interact with one another with respect to cell motility will be essential, and computational modeling will be no doubt an integral part of that endeavor.

## ACKNOWLEDGEMENTS

This work has been supported in part by NIH grants CA 085139 and CA 113004 and by the Harvard-MIT (HST) Athinoula A. Martinos Center for Biomedical Imaging and the Department of Radiology at Massachusetts General Hospital.

## CAPTIONS

**Table 1. Nomenclature, values or ranges of model parameters.** Real parameter values are taken from the literature where necessary.

**Figure 1. Model setup.** (a) Microenvironment setup for the base case: $\lambda$=707.1 $\mu$m, $\alpha$=1, and with a uniform permission distribution. Resistance is initially kept uniform, but is subsequently modified by the cells. (b) Schematic representation of the cancer cell search decision process.

**Figure 2. Boundary states of the search paradigm space.** Representative simulations with 3 different search paradigms $[C_A, C_P, C_R]$: [1, 0, 0] (top), [0, 0, 1] (middle), [0, 1, 0] (bottom). On the left, the final positions of cells after 24 hours of simulation time are shown, together with the final distributions of each of the three tissue qualities. Displayed on the right are the trajectories of the cells with the best (blue) and worst (red) final displacement effectiveness values, from their initial starting positions at time zero. Note that these represent the three boundary states, and do not reveal any synergistic effects of multiple search strategies.

**Figure 3. Displacement effectiveness for the baseline case.** Displacement effectiveness is plotted for various $C_P$ and $C_R$ combinations, where each combination constitutes a unique search paradigm. This is the baseline case: $\lambda$=707.1 $\mu$m, $\alpha$=1, and with a uniform permission distribution (see **Figure 1**). Note that the total coefficient weights sum to unity, and therefore for each $C_P$ and $C_R$ combination, $C_A = 1 - C_P - C_R$ is implied. Areas where no point exists are where this relation is violated. Displayed are the means of five simulations, where the vertical bars represent the standard error of the mean. The color gradient runs from red, which are regions of high displacement effectiveness, to blue, regions of low displacement effectiveness.

**Figure 4. Displacement effectiveness across range of initial cell locations.** Normalized DE for various distances of initial cell position to the chemoattractant source. Displayed are the means of five simulations, where the vertical bars represent the standard error of the mean.

**Figure 5. Displacement effectiveness across range of diffusion coefficients.** Normalized DE for varying diffusion coefficients, where $D^* = \alpha D_{glucose}$. Displayed are the means of five simulations, where the vertical bars represent the standard error of the mean. These plots can be contrasted with the baseline case, $\alpha = 1$, which is shown in Figure 3.

**Figure 6. Displacement effectiveness versus diffusion coefficient for the $C_A$ search paradigm.** Plot of normalized DE versus $\alpha$ for the search paradigm $[C_A, C_P, C_R]$: [1, 0, 0].

**Figure 7. Displacement effectiveness over various peak permission locations.** A permission gradient is introduced to the baseline case (see **Figure 1**), with its peak set to four different locations shown with respect to the initial location of the cells in (a). Displayed are the means of five simulations, where the vertical bars represent the standard error of the mean. Note that the chemoattractant source is located at (75, 75), which corresponds to a distance of +707.1 $\mu$m. Distances of (b) -1414, (c) -707.1, (d) 0, and (e) 707.1 $\mu$m are shown.

**Figure 8. Displacement effectiveness versus tumor heterogeneity.** The Shannon index, which is a measure of diversity of a population, represents the heterogeneity of the cancer cell population (n=100). Diversity here refers to the number of diverse search paradigms taken on by the population of cells at the start of a simulation. The two correlation values are taken from a linear fit and exponential fit of the mean values. Vertical bars represent the standard errors, where 200 independent runs were performed for each point displayed.

# FIGURES & TABLES

## Table 1.

| Parameter | Description | Value or range |
|---|---|---|
| $\Delta t_{diff}$ | Time step for diffusion | 20 s |
| $\Delta t_{cell}$ | Time step for cell migration | 60 min |
| $l$ | Lattice size | 20 μm |
| $N$ | Number of cells | 100 |
| $v_{max}$ | Maximum cell velocity | 20 μm/hr |
| $D_{glucose}$ | Glucose diffusion rate | 67 μm$^2$/s (Sander and Deisboeck, 2002) |
| $U_{glucose}$ | Glucose uptake rate per cell | 0.77 pM/hr (Shrivastava et al., 2006) |
| $S_{glucose}$ | Glucose source | 2.36 nM (Zhang et al., 2007) |
| $C_A$ | Cell chemoattraction coefficient | $0 - 1$, $C_A + C_P + C_R = 1$ |
| $C_P$ | Cell permission coefficient | $0 - 1$, $C_A + C_P + C_R = 1$ |
| $C_R$ | Cell resistance coefficient | $0 - 1$, $C_A + C_P + C_R = 1$ |
| $\delta_{resistance}$ | Reduction in resistance by proteases | 1 |

**Figure 1.**

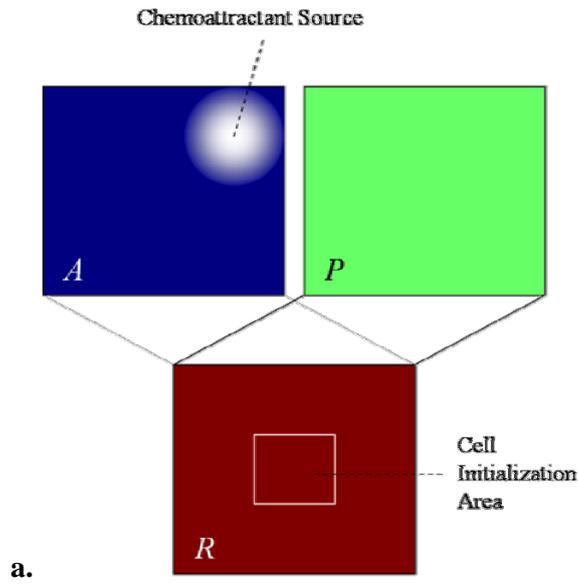

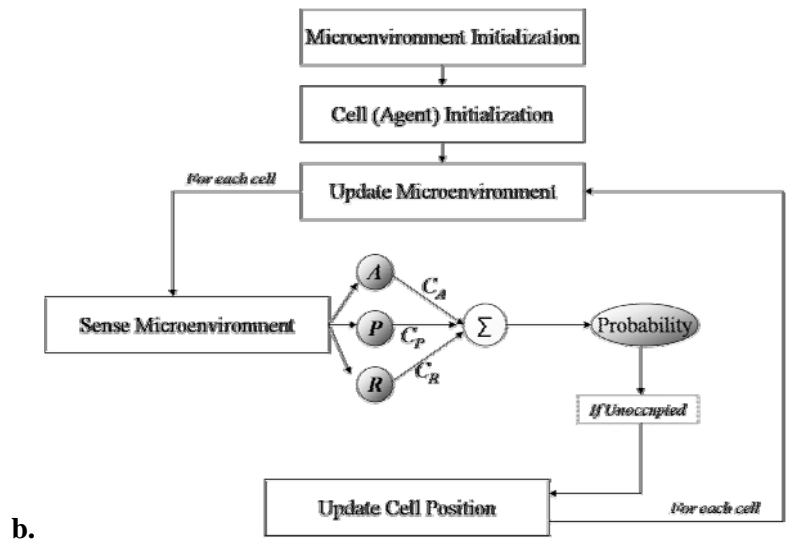

Figure 2.

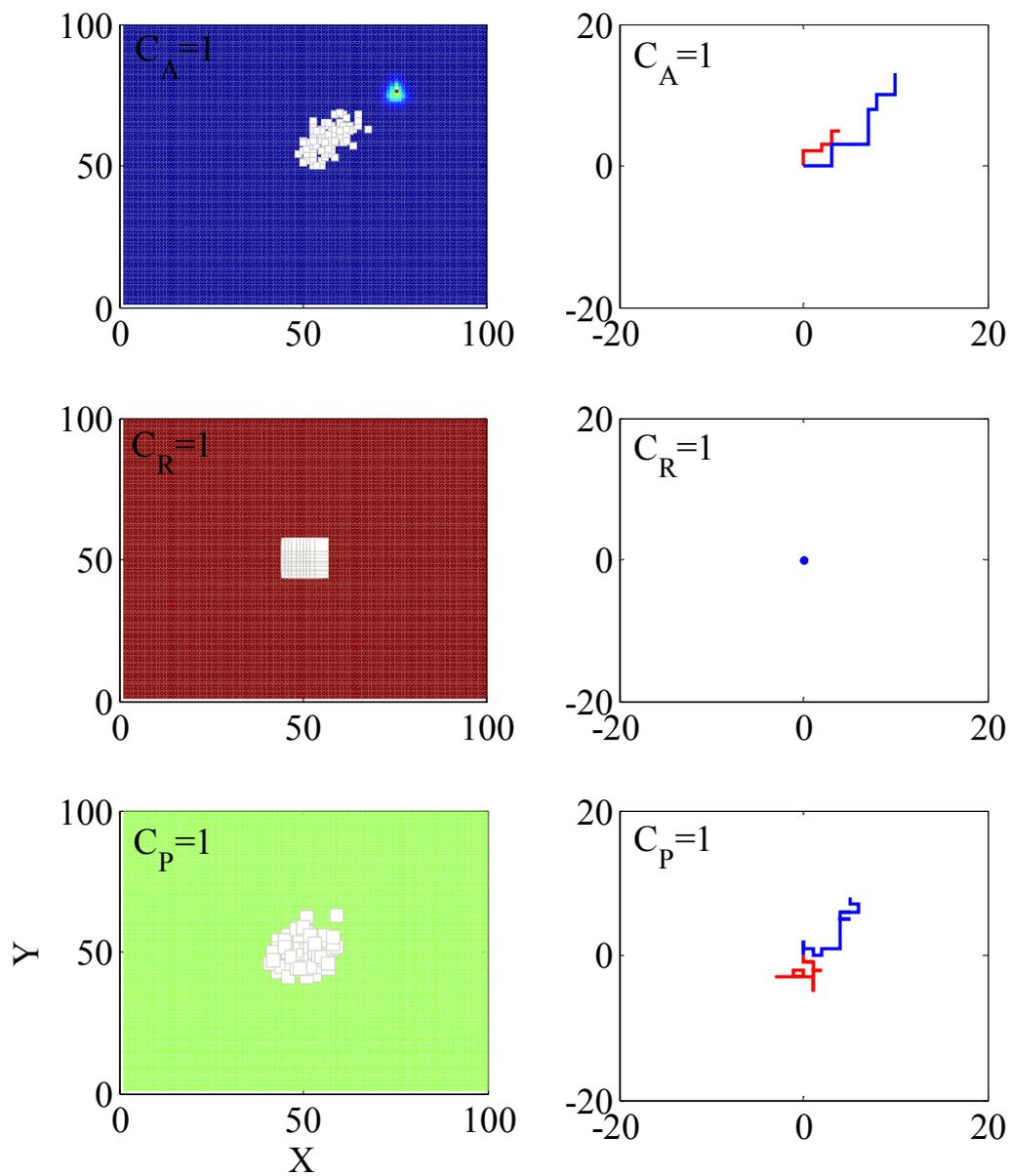

**Figure 3.**

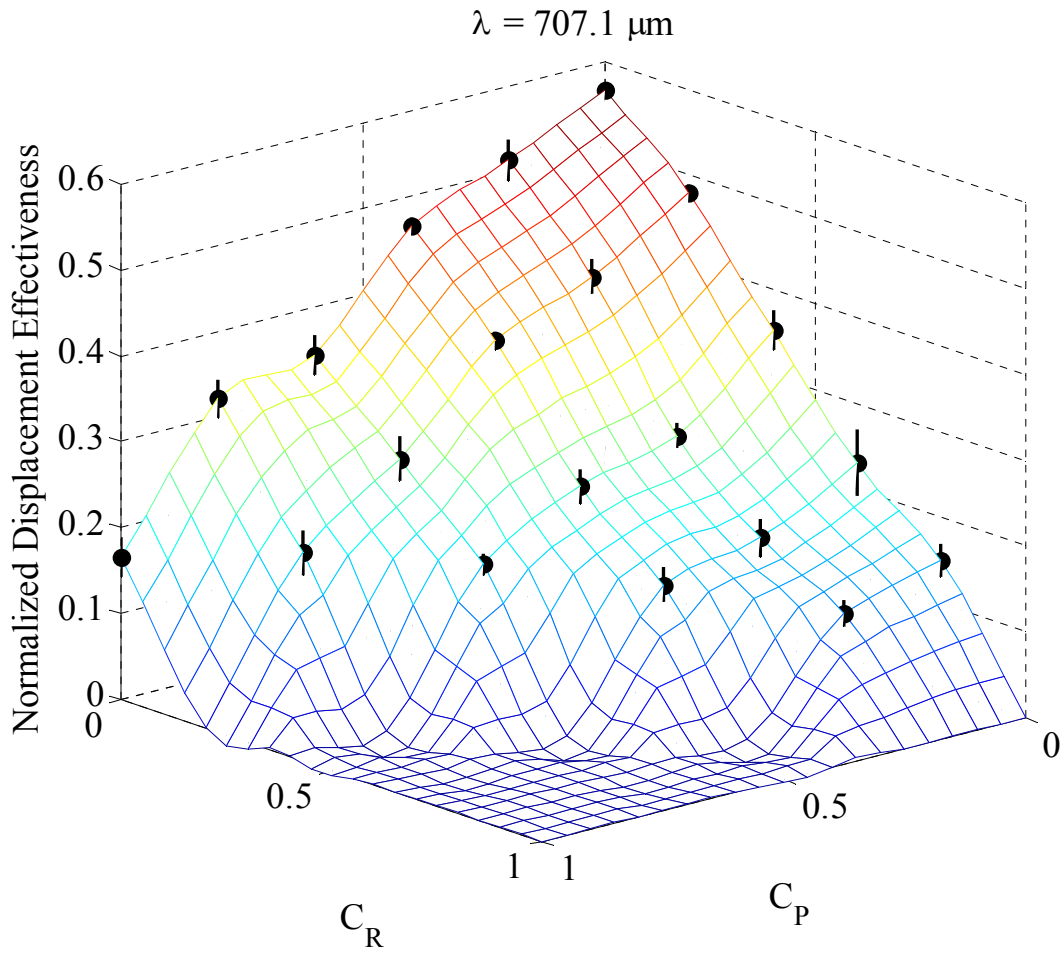

**Figure 4.**

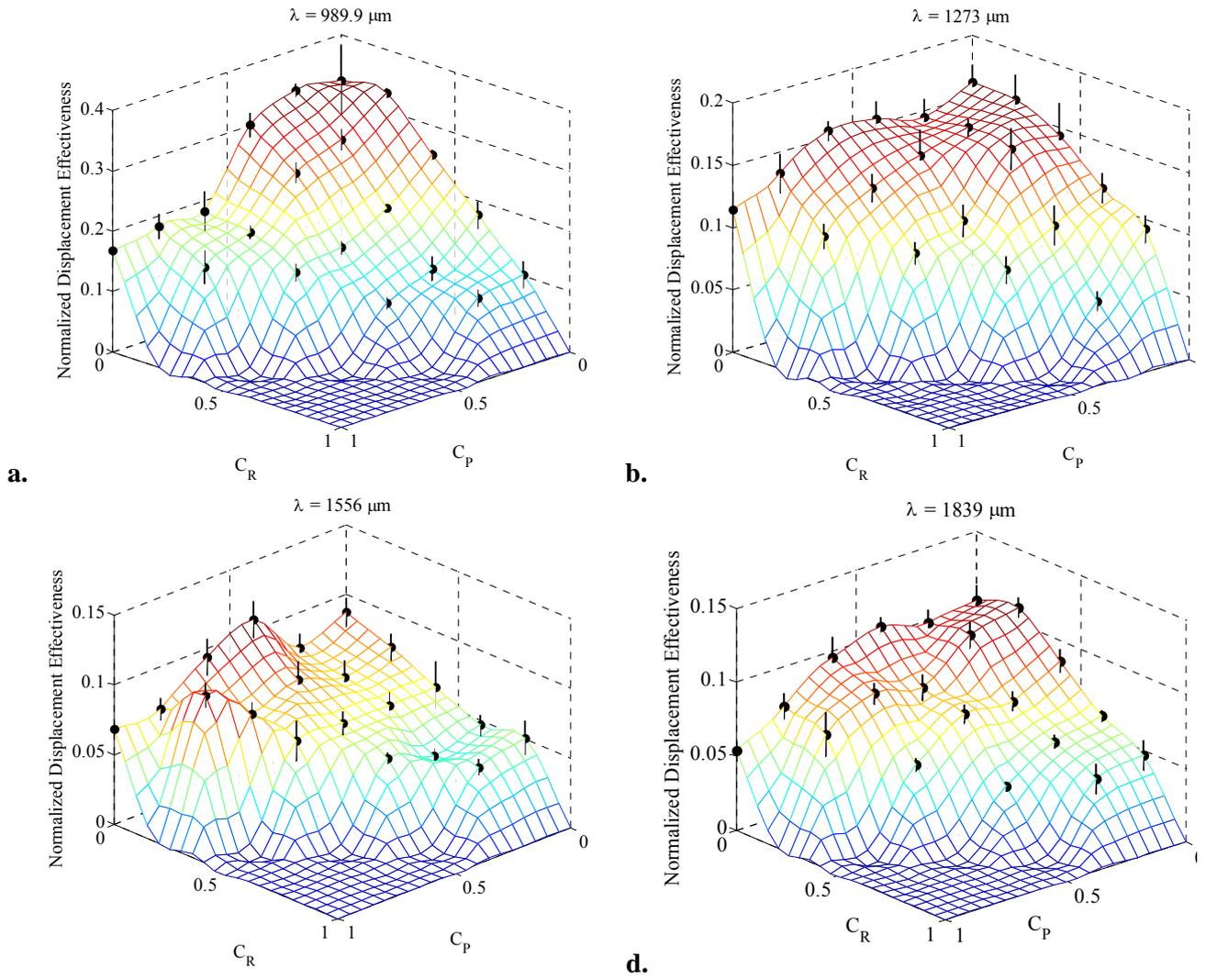

**Figure 5.**

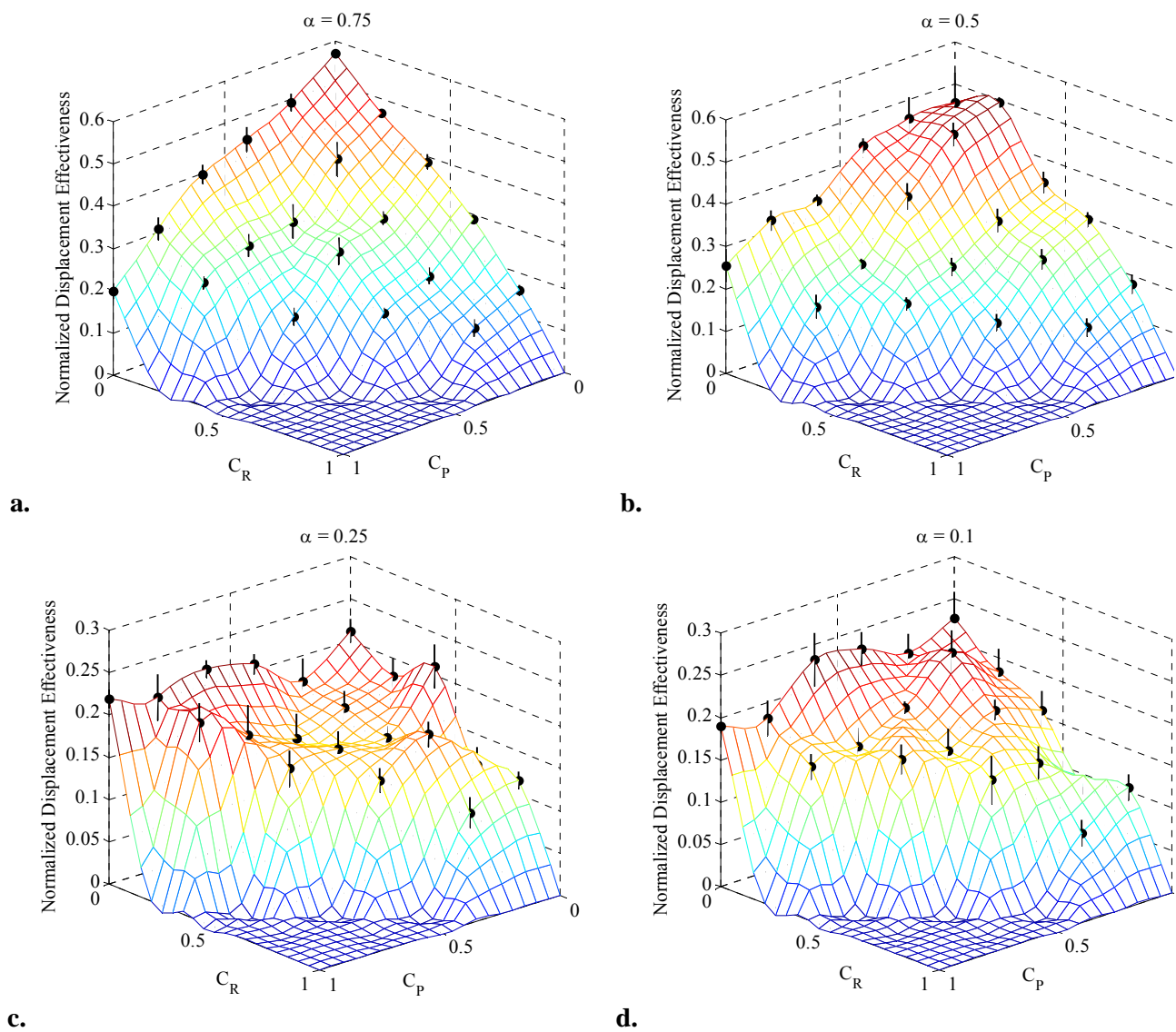

**Figure 6.**

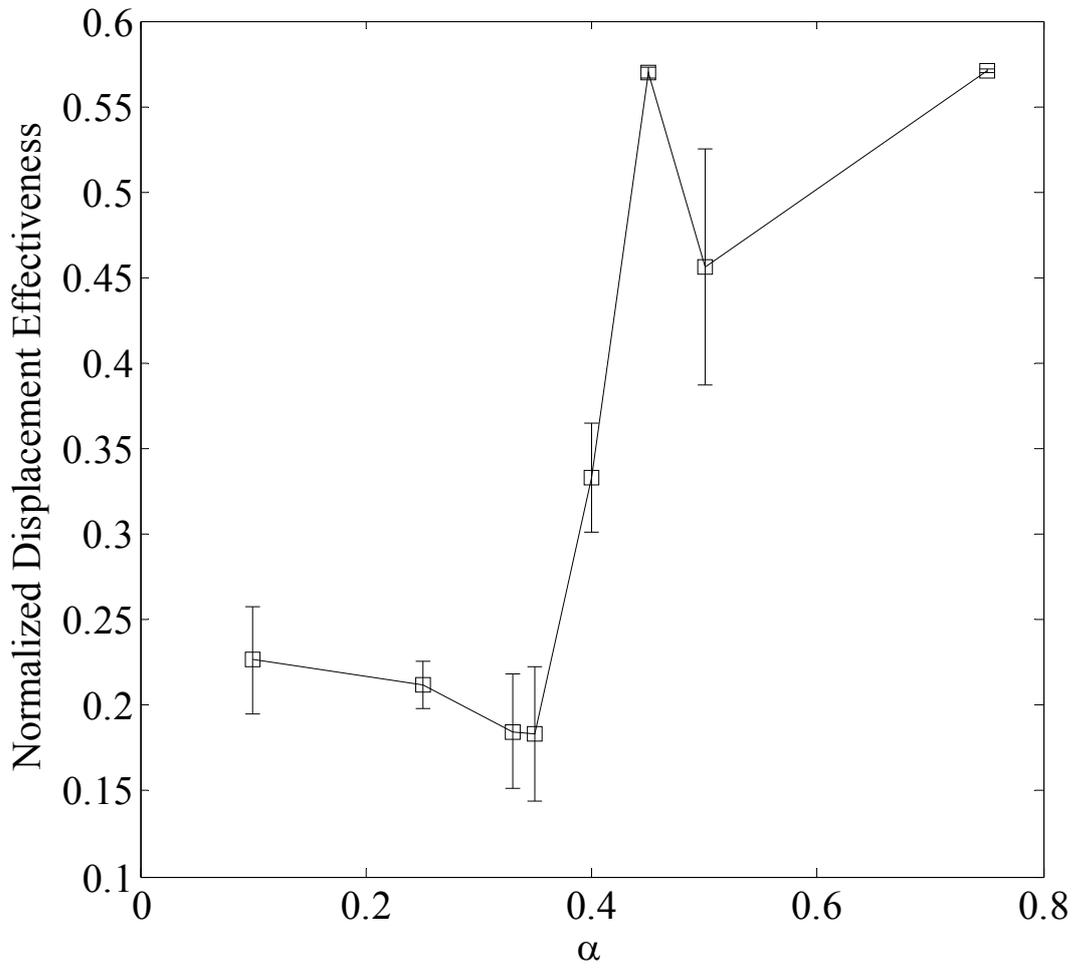

## Figure 7.

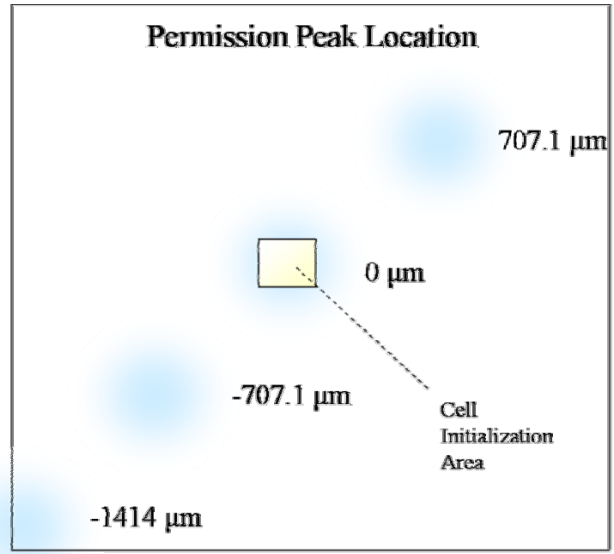

**a.**

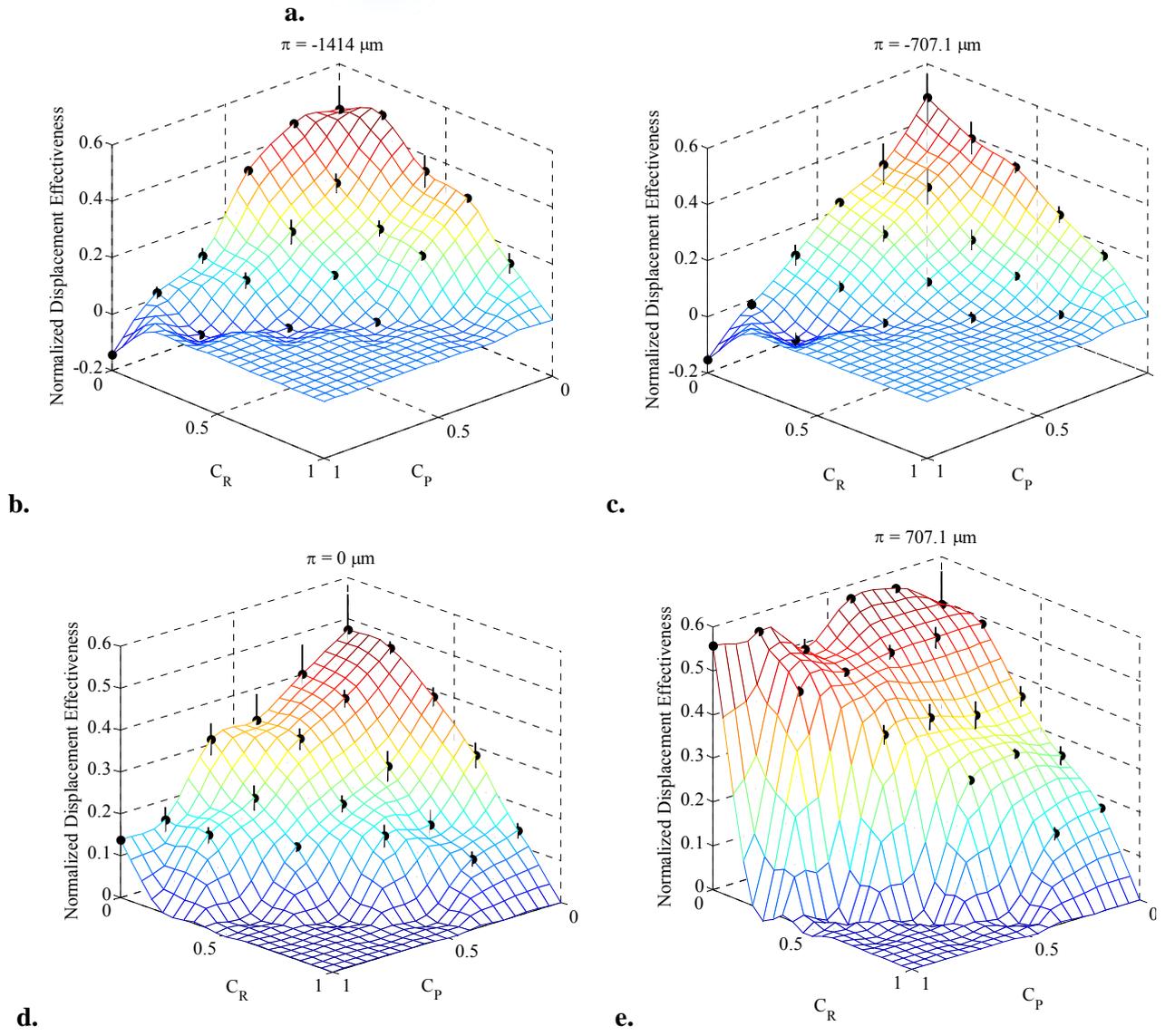

**Figure 8.**

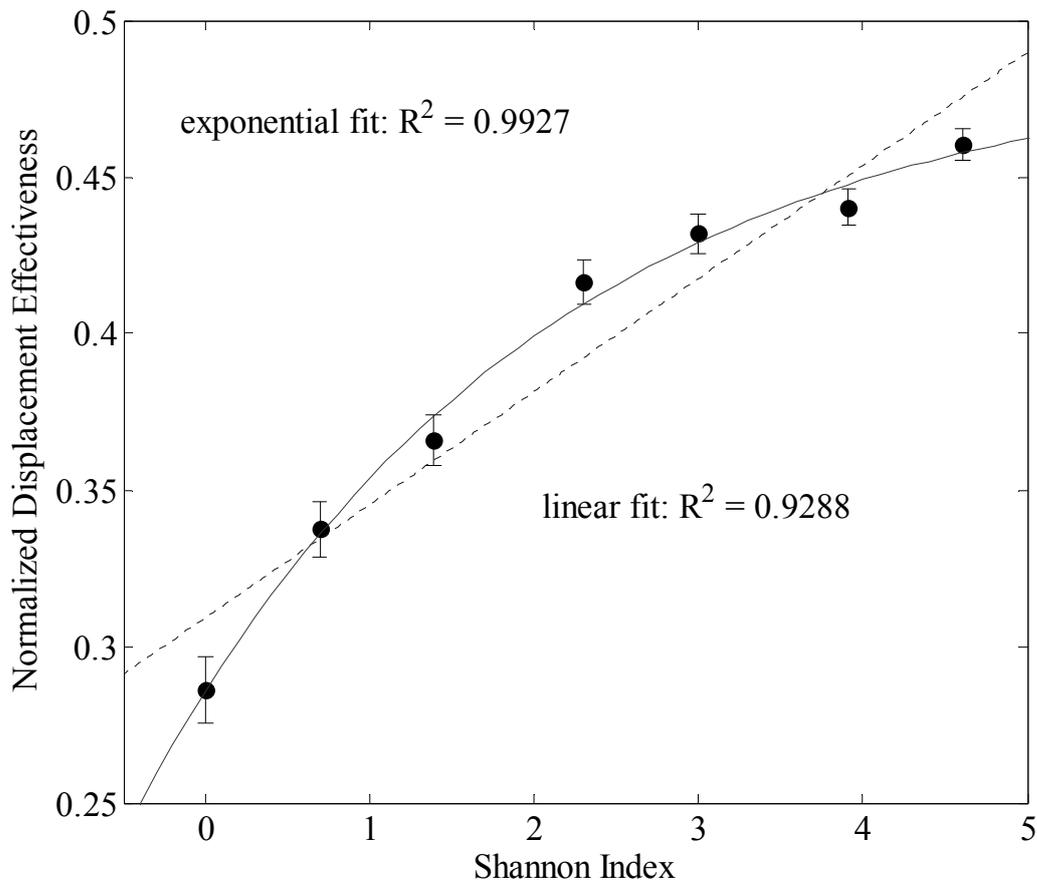